\documentclass{sig-alternate}

\usepackage[caption=false]{subfig}
\usepackage{lipsum}
\usepackage{tabularx}
\usepackage{color}
\usepackage[ruled]{algorithm2e}
\usepackage{graphicx}

\newcounter{HWNumberOfComments}
\stepcounter{HWNumberOfComments}
\newcommand{\hw}[1]{}

\definecolor{DarkGreen}{rgb}{0.000000,0.6,0.000000 }

\DeclareMathOperator*{\argmax}{argmax}

\begin{document}
%
\conferenceinfo{KDD '15}{Sydney, Australia}

\title{Scheduling Broadcasts in a Network of Timelines}
%
%
%
%
%

\numberofauthors{3} 
%
\author{
%
%
\alignauthor
       Emaad Manzoor\titlenote{Research performed while the author was graduate
       student at King Abdullah University of Science and Technology.}\\
       \affaddr{Carnegie Mellon University}\\
       \affaddr{Pittsburgh, USA}\\
       \email{emaad@cmu.edu}
\alignauthor
       Haewoon Kwak\\
       \affaddr{Qatar Computing Research Institute}\\
       \affaddr{Doha, Qatar}\\
       \email{hkwak@qf.org.qa}
\alignauthor
Panos Kalnis\\
       \affaddr{King Abdullah University of Science and Technology}\\
       \affaddr{Thuwal, Saudi Arabia}\\
       \email{panos.kalnis@kaust.edu.sa}
}

\maketitle
\begin{abstract}
Broadcasts and timelines are the primary mechanism of information exchange in online social platforms today. Services like Facebook, Twitter and Instagram have enabled ordinary people to reach large audiences spanning cultures and countries, while their massive popularity has created increasingly competitive marketplaces of attention. Timing broadcasts to capture the attention of such geographically diverse audiences has sparked interest from many startups and social marketing gurus. However, formal study is lacking on both the timing and frequency problems. We study for the first time the broadcast scheduling problem of specifying the timing and frequency of publishing content to maximise the attention received.

We validate and quantify three interacting behavioural phenomena to parametrise social platform users: information overload, bursty circadian rhythms and \textit{monotony aversion}, which is defined here for the first time. We formalise a timeline information exchange process based on these phenomena, and formulate an objective function that quantifies the expected collective attention. We finally present experiments on real data from Twitter, where we discover a counter-intuitive scheduling strategy that outperforms popular heuristics while producing fewer posts.

\end{abstract}




\section{Introduction}

\begin{figure}
	\centering
	\includegraphics[width=0.45\textwidth]{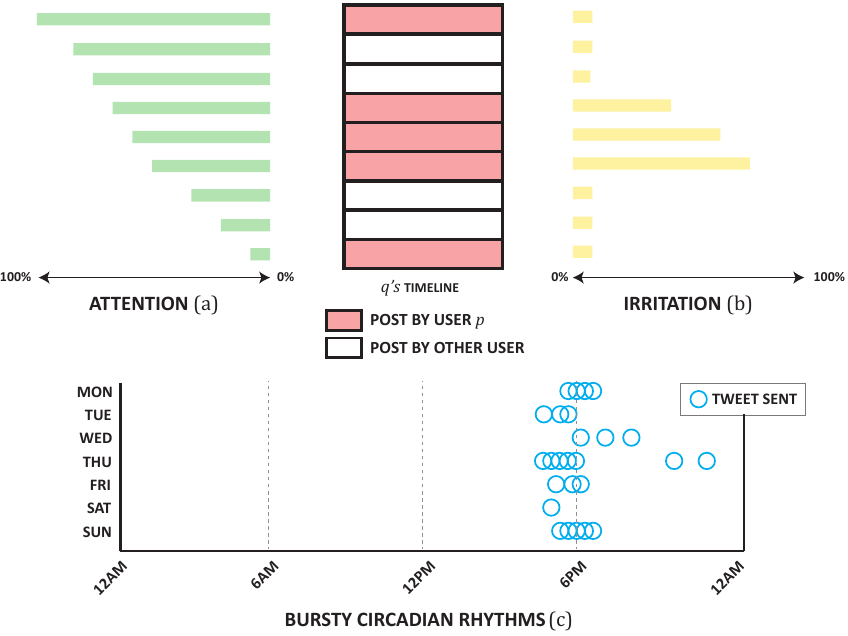}
	\vspace{-2mm}
	\caption{Information overload, monotony aversion and bursty circadian rhythms.}
	\vspace{-2mm}
	\label{fig:introduction-a}
\end{figure}

The advent of social publishing platforms has democratised the attention economy, once a monopoly of radio, television and print media powerhouses. Social platforms have emerged for publishing images (Flickr, Instagram), video (Vine) and a hybrid of both (Facebook, Twitter). A distinctive commonality among these platforms is their mechanism of information exchange, which is primarily via the combination of broadcasts and timelines.

The typical social platform user composes ``posts'', subscribes to posts from other users by ``following'' them, propagates posts to her subscribers by ``resharing'' them, and reacts to posts via ``likes'' or ``favourites''. These actions are instantly broadcast to all her subscribers and appear as a sequence of posts on their timelines. Timelines are traditionally ordered reverse-chronologically as in Twitter, Vine and Instagram. They may also be ordered by a mixture of recency and social metrics, as in Facebook. The timeline is each user's personalised entry point and primary means of engagement with the social platform.

The massive popularity of social platforms has led to the production of large and fast torrents of information. Users, ever in search of novel and interesting content, subscribe to new information inflows without restraint \cite{gomez2014quantifying}. They are soon overwhelmed when their cognitive information processing limits are exceeded by their subscribed inflow, and enter a state of \textit{information overload}. A distinctive symptom of this state is the partial consumption of timelines; posts too low in the timeline often go unseen. With audiences shared across multiple content producers, this results in a marketplace of attention that is growing increasingly competitive.

Concretely, competing content producers vie for the highest slots in each user's timeline. In unicast mediums like email, it is easy to place a message near the top of a target's mailbox by simply sending it just before she checks her email. The scenario in social platforms is challenging for two reasons. The first is user diversity; subscribers may have varying periods of activity on the social platform, observed as \textit{bursty circadian rhythms}. The second is the restriction to broadcasts; a post timing configuration affects \textit{all} subscribers, and must hence strike a balance between them.

Many commercial offerings have sprung up that enable scheduling posts, from established giants like Twitter\footnote{https://archive.today/q1RFb} to startups like Buffer\footnote{https://archive.today/nIQnb} and HootSuite\footnote{https://archive.today/WjHZZ}. A common practice is to visualise subscriber activity as a 2-D histogram across days of the week and allocate posts to the time slots that appear most active. This is lacking in the following ways: it does not specify the post frequency in a time slot, it ignores competing content producers, and it becomes increasingly useless as subscriber diversity increases. Indeed, the strategy makes no attempt at quantifying the ``goodness'' of a timing configuration and serves only as a rough heuristic.

Another possible strategy is to broadcast rapidly in all time slots, flooding the timelines of all subscribers and flushing competitors out. This guarantees attention, but has been observed to irritate users, who may ignore \cite{counts2011taking,comarela2012understanding} or even break ties \cite{kwak2011fragile} with the offending content producer. We attribute this irritation to an increase in the timeline monotony of subscribers who prefer novel content, and term this phenomenon \textit{monotony aversion}.

These three key phenomena: \textit{bursty circadian rhythms}, \textit{monotony aversion} and \textit{information overload} are what necessitate scheduling broadcasts. Their interaction is illustrated in Figure \ref{fig:introduction-a}. Depicted is the timeline of a user $q$ containing posts by user $p$ among others. Also indicated is her variation in (a) attention and (b) irritation as she consumes her timeline from the top downwards. Due to information overload, her attention decreases with timeline depth. She is also irritated by closely-spaced clusters of tweets from the same user $p$. Due to her workday and biological sleep-cycle, her tweeting activity exhibits bursty circadian rhythms (c): she usually tweets at around 6PM, followed by a long period of inactivity while she is at work or asleep.

In this first work on scheduling broadcasts in a network of timelines, we study the problem of allocating a content producer's posts to time slots to maximise the overall attention expected from a given set of subscribers. To quantify the expected attention, we define the \textit{attention potential} of a schedule as a function of (i) the degree of information overload, (ii) the degree of monotony aversion and (iii) the position of the post on the timeline of each subscriber, which depends on the subscribers' and competitors' activity patterns. The link between schedules, activity patterns and obtained attention potentials depends on the specific information exchange process in the network.

A complete formulation of this information exchange process is nontrivial. Timeline construction involves complex temporal interactions between the content producers, subscribers and competitors. Timeline consumption behaviours vary across users with different behavioural characteristics. None of these behaviours can be directly quantified, and inferring attention from observable signals such as retweets, likes and comments tends to be an underestimate \cite{Bernstein:2013:QIA:2470654.2470658}.

Hence, we focus on data-driven analyses of social activity traces to validate the existence of these behaviours and evaluate their significance. This provides both an understanding of the strength and nature of their influence on attention.

In summary, we make the following contributions:

\begin{itemize}
       \item We validate the existence of bursty circadian rhythms in microblogs by studying the distribution of time between consecutive tweets in Sina Weibo. The distribution reveals a pattern that is characteristic of bursty behaviour modulated by circadian rhythms.

       \item We define the concept of monotony aversion and study its impact on attention towards posts. Using data from Twitter, we show that ``closely spaced'' posts on timelines tend to be skipped over inattentively. We also show that posts lower on the timeline can affect the attention attainable by ones above, suggesting that timelines are consumed in chunks rather than as sequential independent posts. This quantification has potentially deep-rooted connections to consumer psychology, recommendation diversity and variety-seeking behaviour, which we detail in the related work section.

       \item Grounded in these observations, we formulate the process of information exchange in timelines and define an objective function that captures the attention potential of a schedule. This function quantifies the relationships observed between attention, information overload, monotony aversion and bursty circadian rhythms. We then formulate the broadcast scheduling problem as a nonlinear integer program and discuss algorithmic approaches to solve it.

       \item We analyse a dataset from Twitter and study the attention potential of some popular heuristic schedules. Using the method of marginal allocation, we discover an alternate schedule that, counter-intuitively, outperforms the heuristics while producing fewer posts. Further analysis reveals an intuitive basis for this schedule, arising from a preference to avoid competition while inducing lesser irritation.
\end{itemize}


\section{Factors Influencing Attention}

It is not immediately obvious why scheduling would benefit a content producer in a social network. Indeed, in the case of social network accounts of news providers, timing seems to be irrelevant. News is traditionally broadcast soon after it breaks, and is actively seeked out and shared across the network. A similar situation is likely in the case of celebrities, with fans constantly on the lookout for personal and professional updates.

Such users are advanced enough in the attention economy to command visibility solely by virtue of their identity, but they are not the focus of our work. Instead, we consider the case of the average social network user: one who is prone to being ignored, unfriended and unfollowed at whim. The average user's place in a networked attention economy is influenced by three key behavioural phenomena, which we detail further in this section.


\begin{figure*}[t]
\centering
    \subfloat[]{
        \includegraphics[width=0.23\textwidth]{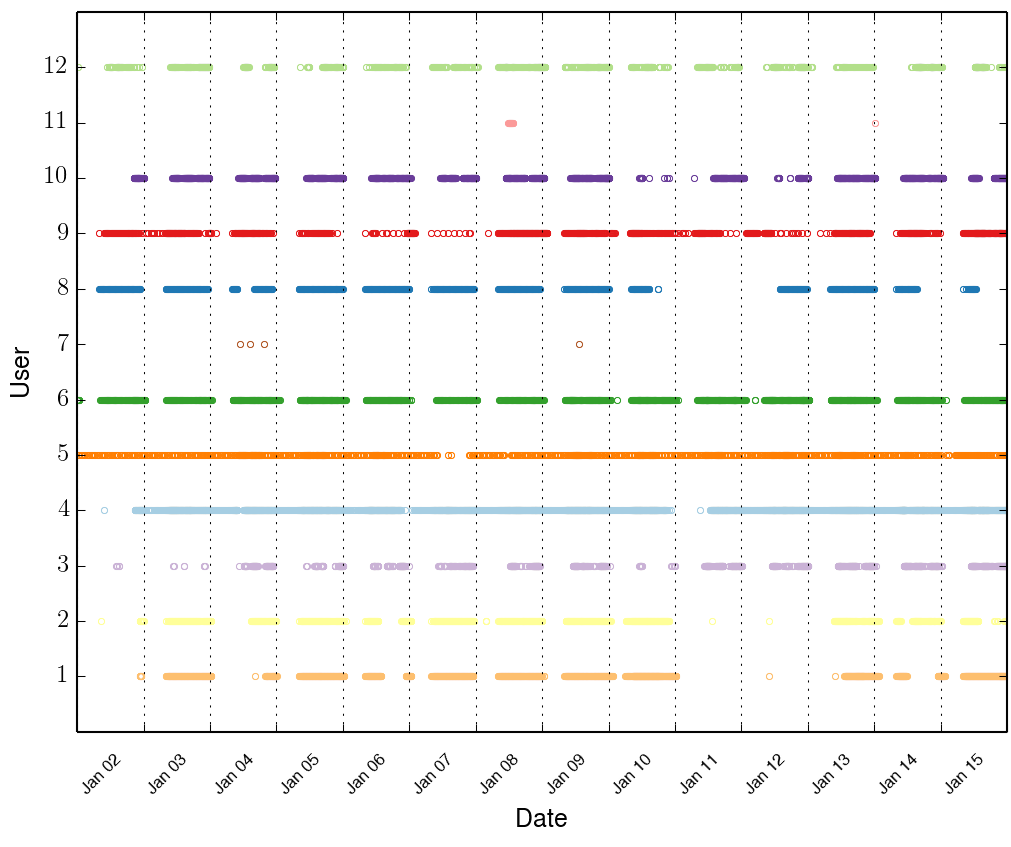}
        \label{fig:weibo-first}}
    \subfloat[]{
        \includegraphics[width=0.23\textwidth]{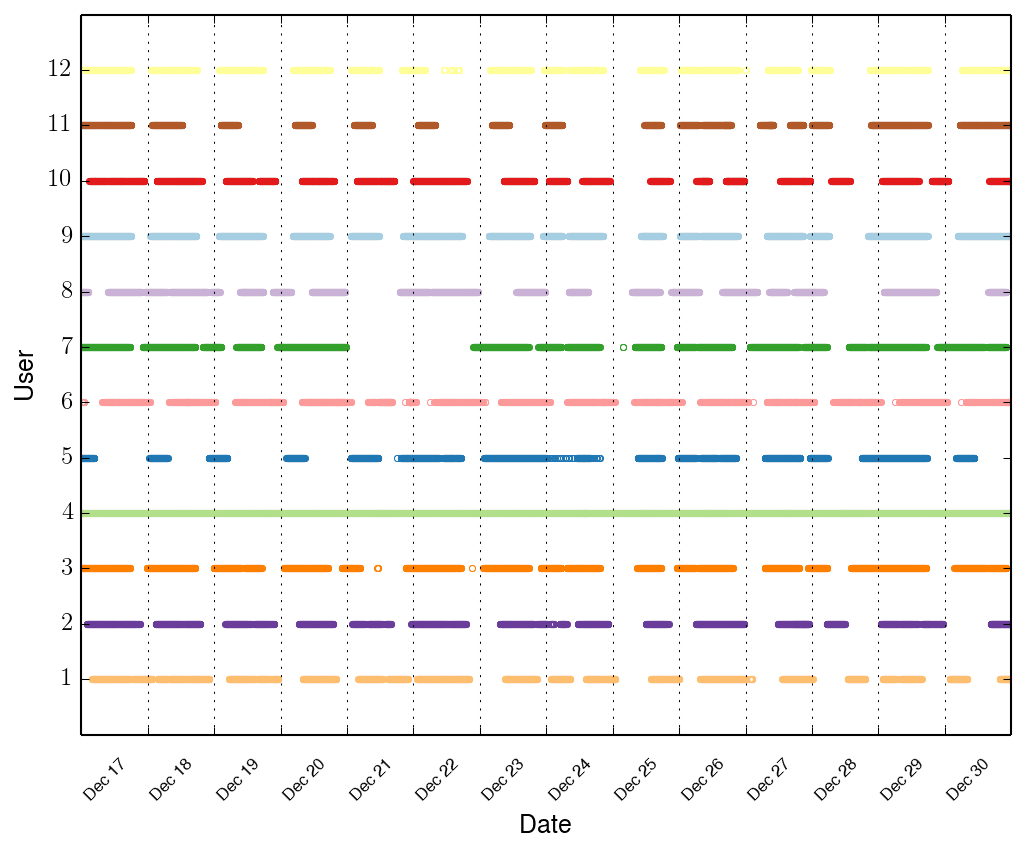}
        \label{fig:weibo-last}}
    \subfloat[]{
        \includegraphics[width=0.24\textwidth]{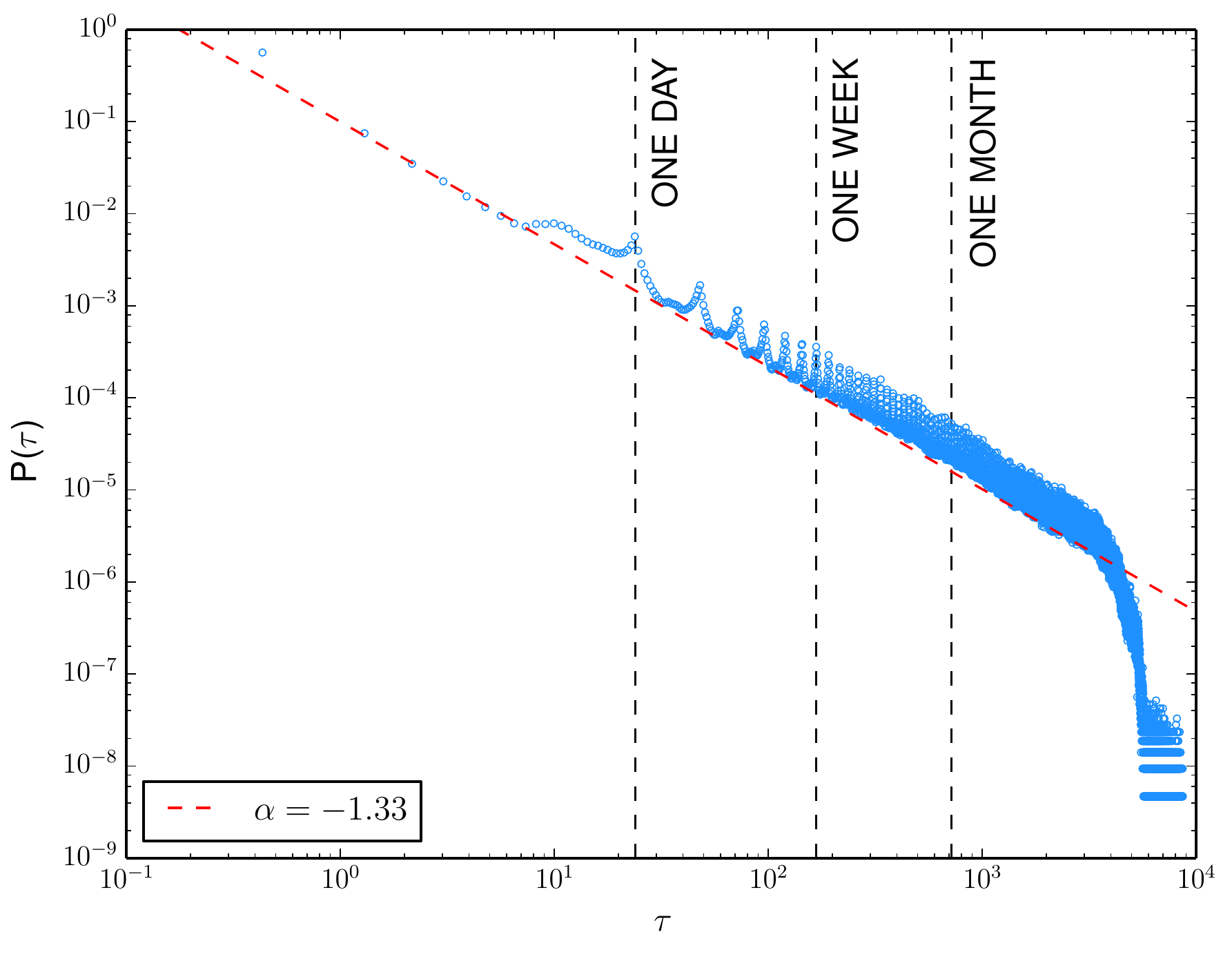}
        \label{fig:weibo-iet}
    }
    \subfloat[]{
        \includegraphics[width=0.24\textwidth]{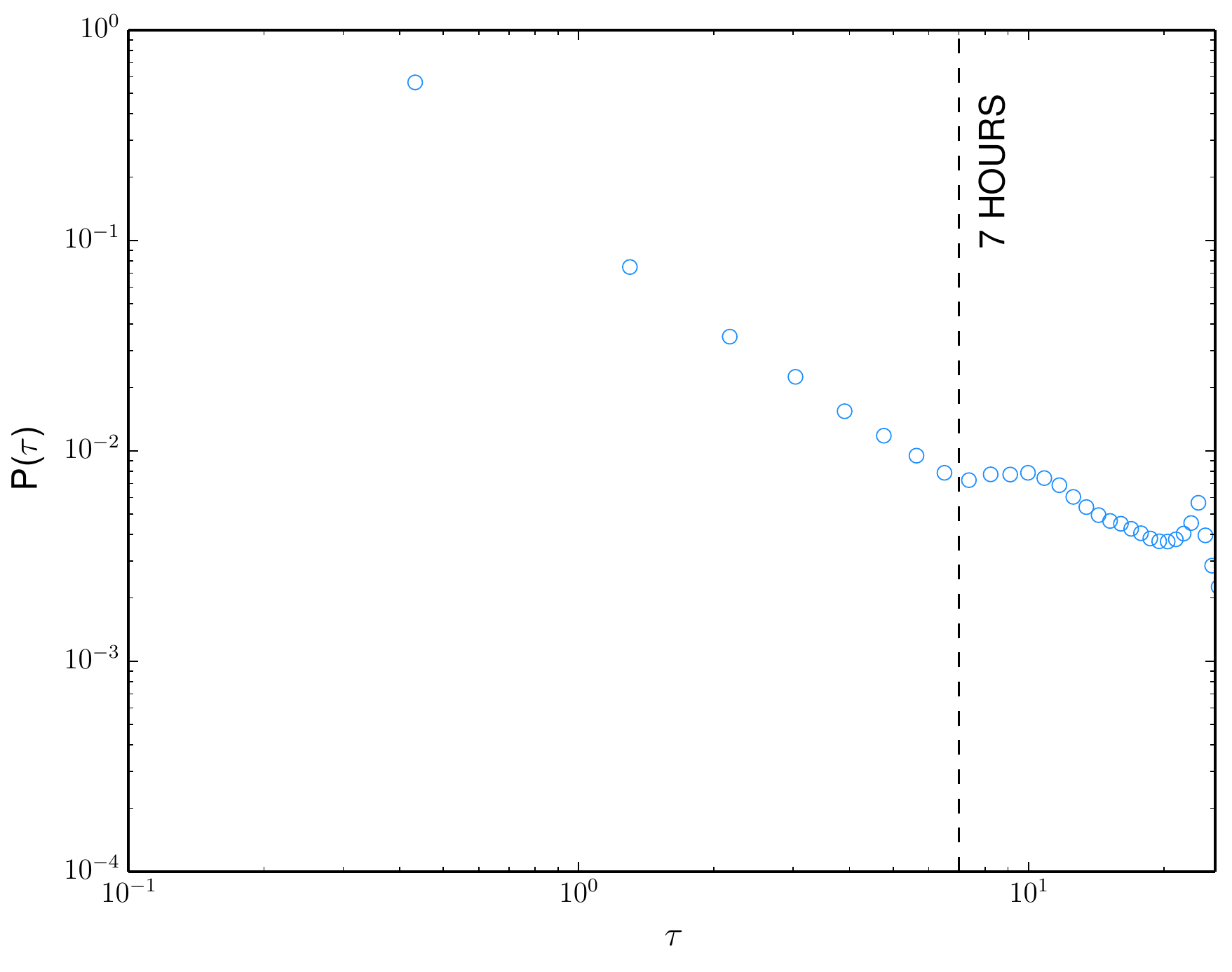}
        \label{fig:weibo-iet-zoomed}
    }
    \caption{Weibo activity for 12 users with the largest number of tweets for the (a) first and (b) last 2 weeks of 2012. Distribution of the time in hours ($\tau$) between consecutive tweets for (c) all $\tau$ and (d) $\tau < 1$ day. }
    \label{fig:weibo}
\end{figure*}

\pagebreak
\subsection{Information Overload}

Information overload has been a subject of intense study in the domains of psychology, sociology and marketing even before the information age \cite{eppler2004concept}. In computer science, it has increasingly been a subject of research on data mining and human-computer interaction \cite{Shahaf:2010:CDN:1835804.1835884}, on recommender systems as choice overload \cite{Bollen:2010:UCO:1864708.1864724}, and on social networks, which we discuss further below.

Information overload arises from exposure to information in excess of one's cognitive processing limit. In social networks, this may be caused by subscribing to either a single overly active user, or to many users whose collective information output causes overload. Hence, research on information overload in social networks has proceeded in two directions. One explores the existence of a cognitive limit on the number of stable social ties that can be maintained at a time and studies how users allocate attention across them \cite{backstrom2011center,gonccalves2011modeling}. The other considers the allocation of attention across items of information. Our interest is in the latter, specifically on attention towards posts on social network timelines.

Oral and eye-tracking surveys \cite{counts2011taking} have reported decreased attention towards information appearing lower on timelines. When timelines are ordered chronologically, it becomes essential to time posts to appear near the top in order to guarantee attention. We do not perform further validation of this phenomenon ourselves, and instead adopt the findings of a recent large-scale analysis of information overload on Twitter \cite{gomez2014quantifying}. Specifically, we note that the  probability of retweeting a tweet decreases monotonically with the tweet's depth in the timeline.


\subsection{Bursty Circadian Rhythms}

Humans have been observed to exhibit bursts while performing sequential tasks in a number of different settings. Such behaviour is characterised by short sessions of high activity followed by long periods of inactivity, giving rise to interevent time distributions that follow a power law. The circadian cycle is also a powerful factor that modulates human behaviour, externally exhibited as rhythms of activity influenced by the biological clock and workday routine. The influence of circadian rhythms have been observed in situations as diverse as diurnal mood changes on Twitter \cite{golder2011diurnal}, patterns of Wikipedia edit activity \cite{yasseri2012circadian} and mobile phone communication \cite{jo2012circadian}.

The combination of bursty activity and circadian rhythms gives rise to interevent time distributions with distinct regularities. The distributions follow a similar power-law curve to that arising from bursty activity, but are characterised by systematic deviations from the curve at 24 hour intervals. This pattern has been observed in a few large-scale studies on online user behaviour, including ones on user activity on a social network \cite{chun2008comparison} and posting behaviour on a blog portal \cite{kim2013microscopic}.

Microblogs differ from traditional content-sharing mediums in their intrinsically real-time nature and hard restrictions on content length. A majority of microblogging is done via smartphones\footnote{https://archive.today/kNNLe}, and users often report events soon after they occur. However, we failed to find prior work probing microblogs for patterns of bursty circadian rhythms. Hence, we look for evidence by exploring the behaviour of users on Sina Weibo, a popular Chinese microblogging platform.

\textbf{Dataset.} We use the Weiboscope dataset \cite{fu2013assessing}, which contains over 13 million tweets posted by over 14 million users spanning the entire year 2012. All users in this dataset had over 1000 followers at the time of data collection, which limits the likelihood of accounts belonging to spammers. \hw{spams?} In comparison with the literature mentioned earlier in this section, this is a relatively large and temporally extensive dataset. Hence, we expect our findings to reasonably reflect general trends of human behaviour on microblogs. \hw{http://\\www.sciencemag.org/content/333/6051/1878.short this is not directly related with this, but maybe worth to mention.} \hw{ and my Cyworld paper ``Comparison of online social relations in volume vs interaction: A case study of Cyworld''}

To gain some initial intuition, we visualize the tweeting activity of the top 12 (arbitrarily chosen) most active users \hw{any specific reason for choosing 12?} for the first (Figure \ref{fig:weibo}(a)) and last (Figure \ref{fig:weibo}(b)) two weeks of 2012. We observe three broad activity patterns. A few users appear to be inactive, while a few tweet throughout the day. For the remaining majority, bursty behaviour is evident with sessions of activity preceding long intervals of inactivity. The bursts also appear to be periodic, with inactivity recurring during the same intervals of time every day.

We now look for the characteristic patterns arising from the combination of bursty activity and circadian rhythms. Figure \ref{fig:weibo}(c) depicts the distribution of the time $\tau$ between a user's consecutive tweets for all users in the dataset. The plot follows a power-law, $P(\tau) = \tau^{-1.33}$ (estimated using \textsc{powerlaw} \cite{alstott2014powerlaw}), eventually terminating in a skewed curve with a heavy tail. The periodic deviations from the power law curve are distinctly visible. \hw{can we get the alpha ($y=x^\alpha$) of this distribution?} If we zoom in on the curve in the region $\tau < 1$ day (Figure \ref{fig:weibo}(d)), we notice a pair of local optima at $\tau = 7$ and $\tau = 10$ hours. These can be explained by considering the average workday that is 8-10 hours long; users are more likely to resume tweeting after work, leading to the local maximum after the minimum at $\tau = 7$ hours. \hw{can we change the unit of x-axis of Figure 7 from seconds to hours? it makes readers easily follow this paragraph.}

Prior works have studied and compared these distributions across many different scenarios, mostly arriving at similar conclusions on the behavioural phenomena that generate them. The aforementioned local optima were also observed in an analysis of multiplayer gaming activity \cite{mryglod2015interevent}, similarly hypothesized as a result of the average workday. In a study on inter-visit time in the Korean social network Cyworld \cite{chun2008comparison}, the distribution curve is further dissected into three power-law regimes, and the phenomena giving rise to each of them are hypothesized and compared to other settings. For our purpose, it suffices to find evidence that there exists an interval of time when the user is inactive, during which microblog posts from her followees will accumulate on her timeline without being seen. The observations discussed in this section strongly indicate the existence of such intervals.

\subsection{Monotony Aversion}

\begin{figure}[t]
\centering
\includegraphics[width=0.45\textwidth]{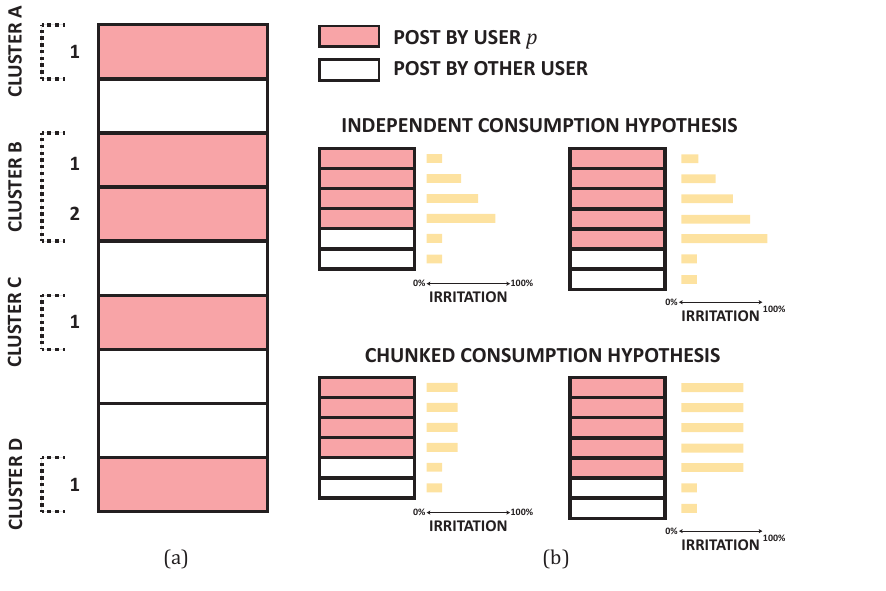}
\vspace{-2mm}
\caption{(a) Clusters and the cluster positions of each producer tweet and (b) the two hypotheses of monotony aversion.}
\label{fig:monotony-a}
\vspace{-2mm}
\end{figure}

To maximise visibility in the face of limited attention in online social networks, a naive strategy is to rapidly broadcast, completely filling every follower's timeline. This raises the concern of irritation on viewing a monotonous timeline dominated by a single user's posts. The phenomenon of users exhibiting irritation with monotony, which we term \textit{monotony aversion}, makes brief appearances in a number of scenarios.

An eye-tracking study of Twitter users viewing their timelines \cite{counts2011taking} reports decreased attention towards users whose tweets appear more than 5 times in a short viewing window. Facebook users too report similar feelings \cite{koroleva2011cognition}, stating that high posting frequencies tend to irritate them, especially when originating from weaker ties. At its extreme, irritation may result in breaking ties with the offender, which terminates the inflow of information from that user \textit{regardless of the tweet content} \cite{kwak2011fragile}.

We thus suspect closely-spaced posts on a user's timeline to irritate her by reducing her timeline's diversity. Now we study the mechanisms behind this phenomenon by observing how users' reactions vary across tweets in different spacing configurations on the timeline.

\textbf{Definitions.} We define a \textit{cluster} as a group of consecutive tweets on a timeline having the same author, termed the \textit{cluster author}. The \textit{cluster size} is then the number of tweets in a cluster. The \textit{cluster position} of a tweet is its distance from the top of the cluster downwards, such that the chronologically newest tweet is at cluster position 1. These definitions are illustrated in Figure \ref{fig:monotony-a}(a), which depicts the clusters A, B, C and D of size 1, 2, 1 and 1 respectively, and the cluster position of each tweet by $p$.

Since a tweet that is retweeted or replied to must have been viewed, we use these \textit{reactions} to tweets as a proxy for attention. This is a conservative measure because users generally react to only a small fraction of the tweets that they view. This is evident in the small reaction probabilities that show up in our forthcoming experiments.

\begin{table}[t]
\centering
\begin{tabular}{| l | l | l |}
    \hline
    \textbf{Cluster Size} & \textbf{\# Reactions} & \textbf{\# Total Tweets} \\
    \hline
    1 &  15897 &   8435832 \\
    2 &  2756 &    1819014 \\
    3 &  710 & 586665 \\
    4 &  304 & 243536 \\
    5 &  126 & 126125 \\
    6 &  79 &  72486 \\
    7 &  54 &  49119 \\
    8 &  21 &  26376 \\
    9 &  16 &  17019 \\
    10 & 15 &  13600 \\
    >10 & 28 & 49673 \\
    \hline
\end{tabular}
\caption{\textit{Twitter-Friends} tweet cluster statistics.}
\label{table:cluster-statistics}
\end{table}

\textbf{Dataset.} We use the \textit{Twitter-Friends} dataset \cite{lin2014steering}, consisting of a follow-graph induced by 822 users with 56,286 links and all the tweets posted by them in the year 2011. The network is dense, with most users being current or former Twitter employees and active Twitter users. We deliberately restrict our attention to regular individuals whose tweets may or may not receive attention. In contrast, tweets by special users such as celebrities and news channels are actively seeked out by fans and would benefit little from scheduling. Since each tweet is timestamped, we can construct the entire timeline for any user by aggregating and reverse-chronologically  sorting all the tweets by the users she follows. Hence, for any tweet on a user's timeline, we can calculate its cluster size and cluster position by looking at the authors of its neighbouring tweets. Since the graph may have evolved during 2011, our timeline reconstruction is noisy and potentially inaccurate. Table \ref{table:cluster-statistics} displays some statistics of the tweet clusters in this dataset.

We want to observe if and how the cluster size and cluster position of a tweet influence its chances of being reacted to. Let $R \in \{True, False\}$ be the event of a reaction and $C$ the cluster size. We first focus on the empirical probability of a tweet obtaining a reaction given its cluster size, $P(R | C)$, depicted by the dotted line in Figure \ref{fig:retweet-distribution}. We can see that the chances of a reaction reduce as the cluster size increases, suggesting that to obtain more retweets or replies, it is wise to enforce a delay between consecutive tweets. This will allow other users to intersperse their tweets between yours and keep the cluster size low.

\begin{figure*}
\centering
    \subfloat[]{
        \begin{tabular}{| c | c |c | c | c | c |}
            \hline
            & $j = 1$  & $j = 2$ & $j = 3$ & $j = 4$  & $j = 5$\\
            \hline
            $i = 1$  & - & 0.0004 & 0.0007 & 0.0006 & 0.0009 \\
            $i = 2$  & - & -     & 0.0003 & 0.0003 & 0.0005 \\
            $i = 3$  & - & -     & -     & $-5.7\times10^{-5}$ & 0.0002 \\
            $i = 4$  & - & -     & -     & - & 0.0003 \\
            \hline
        \end{tabular}
        \label{table:observed-test-statistic}
    }
\quad
    \subfloat[]{
        \centering
        \begin{tabular}{| c | c |c | c | c | c |}
            \hline
            & $j = 1$  & $j = 2$ & $j = 3$ & $j = 4$  & $j = 5$\\
            \hline
            $i = 1$  & - & \textbf{0.001} & \textbf{0.001} & \textbf{0.001} & \textbf{0.001} \\
            $i = 2$  & - & -     & \textbf{0.001} & \textbf{0.019} & \textbf{0.001} \\
            $i = 3$  & - & -     & -     & 0.669 & 0.109 \\
            $i = 4$  & - & -     & -     & - & 0.077 \\
            \hline
        \end{tabular}
        \label{table:cluster-pvalues}
    }
    \vspace{-2mm}
    \caption{The observed test statistic (a) and p-values for the null hypothesis (b). Cell $(i,j)$ contains (a) $T_{obs}(i,j) = $ $P(R | C = i) - P(R |C = j)$ and (b) the probability that $T_{obs}(i,j)$ is due to chance, where $R$ is the event of a reaction and $C$ is the cluster size. Bold probabilities are significant ($p < 0.05$).}
    \vspace{-2mm}
    \label{fig:tables}
\end{figure*}

However, it is possible that this decrease in reactions is simply due to the relative rarity of larger clusters. Hence, we conduct a statistical randomisation test that indicates with high confidence that the decrease in reaction cannot be due to chance. The observed test statistic is shown in Table \ref{fig:tables}(a). We generate 1000 random permutations of tweets across cluster sizes to construct our null models and compute p-values, shown in Table \ref{fig:tables}(b). We can thus be reasonably confident that $P(R | C = 1) > P(R | C = 2) > P(R | C = 3)$. However, when comparing reaction probabilities between clusters of size 3, 4 and 5, we can no longer reject the null hypothesis. This may explain the noisy disruption in trends for $C > 3$ in Figure \ref{fig:retweet-distribution}. However, clusters with $C > 3$ give rise to less than 0.5\% of all tweets on the timelines in this dataset, so our hypothesis holds for the large majority that remains. A potential root cause of large clusters in our data is noise; tweets missing in the data could have broken large clusters into smaller ones had they been present.

\begin{figure}
    \includegraphics[width=0.45\textwidth]{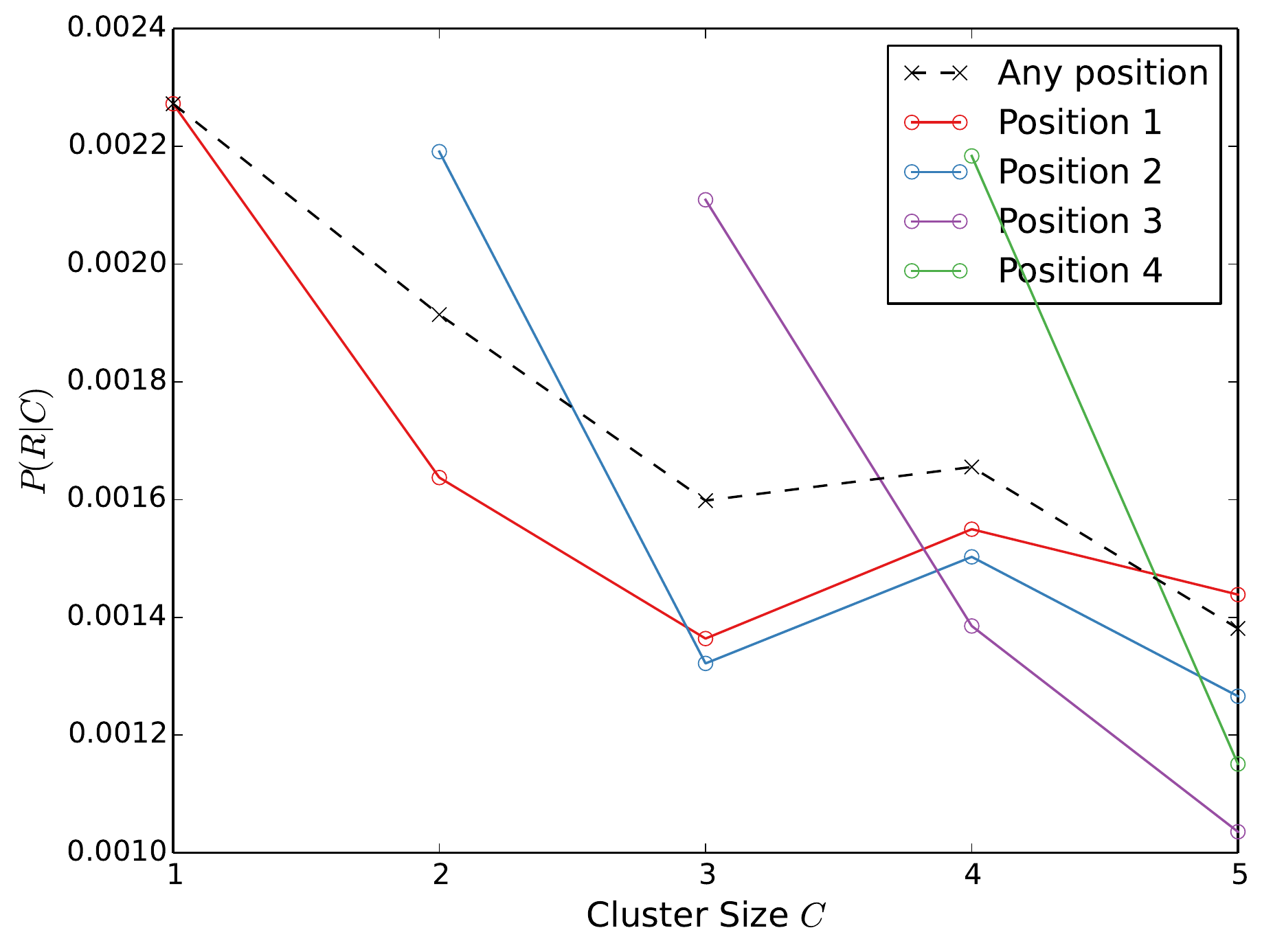}
    \vspace{-2mm}
    \caption{The probability of being reacted to given the cluster size. The dotted line shows this probability across all cluster positions, while the solid lines show this probability given a fixed cluster position.}
    \vspace{-2mm}
    \label{fig:retweet-distribution}
\end{figure}

We now focus on the following question: are the chances of a reaction affected by the number of tweets below it in the same cluster? We would not expect this if users consume tweets sequentially and independently, which we term the \textit{independent consumption hypothesis}. However, the solid lines in Figure \ref{fig:retweet-distribution} indicate that, given a fixed cluster position, reaction probability decreases with increasing cluster size. This supports the alternative \textit{chunked consumption hypothesis}, illustrated in Figure \ref{fig:monotony-a}(a). Specifically, one cannot tweet creating large clusters and still expect the newest tweets in that cluster to command the same attention; there is a reasonable chance of the user skipping over the entire cluster of tweets in irritation.

The results above quantitatively validate the presence of irritation and loss of attention reported by earlier qualitative studies. They reveal the trade-off that a broadcaster must make between creating large and visible but irritating clusters, and small, less visible clusters that command more attention. While our results suggest that the irritated user inattentively skims over offending tweet clusters, large tweet bursts have been reported \cite{kwak2011fragile} as the most common reason to unfollow someone on Twitter. Hence, it is essential for a broadcast scheduling algorithm to intelligently control the perceived spacing of tweets on followers' timelines.


\section{The Broadcast Scheduling\\Problem}

\hw{we need to mention how our findings in previous section connects to this section.  Currently, sections do not connect well.}
In the previous section, we found that users in microblogs exhibit bursty circadian rhythms. We also found that users exhibit monotony aversion, which manifests itself as reduced attention towards large clusters of posts on timelines. We know that users exhibit information overload, only partially consuming their timelines. These phenomena interact to give rise to the broadcast scheduling problem, which we detail further in this section.

The interplay between the actions of the following entities in the social network collectively leads to the construction of each follower's timeline:

\begin{enumerate}
       \item The producer, whose schedule we wish to optimise.
       \item Followers, who follow the producer.
       \item Competitors, who are followed by the followers.
\end{enumerate}

The producer has partial control over each follower's timeline and, together with the competitors, constructs the follower's timeline as a reverse-chronologically ordered sequence of posts. The restriction to broadcasts enforces having a single schedule that impacts all followers. Hence, this schedule must maximise attention by striking a balance between followers with different competitors, circadian rhythms and degrees of information overload and monotony aversion.

To formalise the interactions between the producer, followers and competitors, we begin this section by describing the timeline information exchange process. This links the locations of posts on timelines to the behaviour of the producer, followers and competitors. We introduce behavioural parameters for each follower quantifying their circadian rhythms, monotony aversion and information overload. We then introduce \textit{survival functions} for each follower and post, and relate them to the notion of the \textit{attention potential} of a post. We finally arrive at an objective function quantifying the \textit{attention potential of a schedule}, which we characterise and discuss algorithmic approaches to optimise.

\subsection{Timeline Information Exchange}

\hw{`notations' or something like that is more appropriate section title} We first focus on the behaviour of the producer who desires a schedule specifying how many posts to broadcast at each instant of the day. We assume this schedule is repeated every day. This assumption is simplifying but not restrictive, since our formulation is valid for schedules that stretch over periods of weeks or months as long as they recur after this period. If we discretise a day into $S$ time slots, a broadcast schedule is a set $\mathbf{X} = \{x_0, \dots, x_{S-1}\}$, where $x_i \geq 0$ is the number of posts broadcast in time slot $i$. Let $N$ be the number of posts intended to be broadcast each day. It may not be optimal to broadcast all of one's intended posts, so $\sum_{i=0}^{S-1} x_i \leq N$.

Competitors are considered separately for each follower, since each follower may follow a different set of users. We count the total number of posts produced by all the competitors for a given follower in a time slot. For time slot $i$ and follower $j$, we denote this number by $c_{ij}$, and we have an aggregate competitors schedule $\mathbf{C_j} = \{c_{0j}, \dots, c_{(S-1)j}\}$.

We consider the timeline construction process of a follower $j$ to happen as follows: in each time slot $i$, the producer first broadcasts $x_i$ posts, then the competitors collectively broadcast $c_{ij}$ posts that appear above the producer's posts in the timeline by virtue of being more recent. The broadcasts repeat similarly for slot $i+1$, eventually wrapping around to slot 0 at the start of the next day. The process continues indefinitely, leading to an infinite reverse-chronological sequence of posts by the producer and competitors. This sequence is the follower's timeline.

\begin{figure}
\centering
       \includegraphics[width=0.45\textwidth]{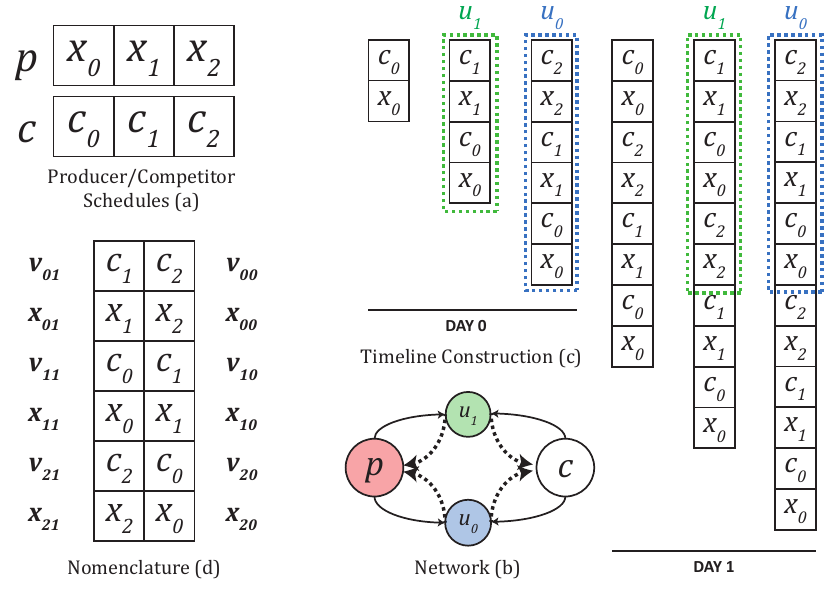}
       \vspace{-2mm}
       \caption{Timeline Construction, Consumption and Nomenclature.}
       \vspace{-2mm}
       \label{fig:timeline-construction}
\end{figure}

Figure \ref{fig:timeline-construction} illustrates this process. Depicted is a network (b) with a single competitor $c$ with 2 followers $u_0$ and $u_1$. There are 3 slots time slots ($S = 3$). The producer schedule $\mathbf{X}$ and competitor schedule $\mathbf{C}$ are depicted in (a). Since both followers have the same competitors, we drop the subscript $j$ for the competitor schedule, and the timeline construction process (c) is common to both.

We now focus on the behaviour of the followers. We assume that followers login to the social platform and consume their timelines once in a day. Concretely, follower $j$ logs in at the end of time slot $\sigma_j \in \{0, \dots, S-1\}$ and begins consuming her timeline from the top downwards, eventually \textit{quitting} the timeline by stopping consumption after reaching some specific \textit{depth}.\hw{or consumes $k$ tweets, where $k$ is randomly drawn from a specific distribution}

In the case of Twitter, a graphical marker denotes the depth after which there are posts that have already been viewed. The post right after marker was, in fact, the first post on the timeline when the follower logged in previously. Hence, we assume that the follower never scrolls past this marker, though she may quit the timeline much before reaching this far.

The order and frequency of posts that the follower views on her timeline remains the same on each login (on average, since each $\mathbf{C_j}$ is an estimate from historical observations). These posts appear in groups called \textit{clusters}, placed on the timeline by the producer and competitors in an interleaved manner. Let there be $x_{ij}$ producer posts in the cluster at position $i$ (from the top) for follower $j$, and $v_{ij}$ competitor posts right above it. We will relate the number of posts in a cluster with the producer and aggregate competitor schedules later in this section.

Figure \ref{fig:timeline-construction} illustrates this nomenclature and the timeline consumption process. Followers $u_0$ and $u_1$ log in at time slot 2 and 1 respectively ($\sigma_0 = 2, \sigma_1 = 1$). On day 0, $u_1$ consumes her timeline (marked in green) in time slot 1 for the first time, which stabilises on day 2 and remains the same every day henceforth. Similarly, $u_0$ consumes her timeline (marked in blue) in time slot 2. The timelines viewed by each follower on every login on are depicted in (d), with the number of posts in each cluster labelled according to our nomenclature.

Having formalised the process governing the position of each post in relation to the producer and aggregate competitor schedules, we can now continue towards defining the \textit{attention potential} of each post on a timeline.

\subsection{Attention Potential}

The extent to which a follower scrolls down the timeline depends on her degree of information overload. We define a parameter $\rho_j \in [0,1]$ for each follower $j$ that captures her intrinsic tendency to quit consuming the timeline. $\rho_j \in \{0,1\}$ correspond to full and no timeline consumption, respectively. We consider a follower as having \textit{survived} until depth $d$ if she scrolls down at least until the $d^{th}$ post. We define a \textit{follower survival function} $F(d;\rho_j)$ that quantifies the probability of the follower $j$ surviving until depth $d$.

Similarly, we also discuss the notion of \textit{cluster survival}. We consider a cluster of posts as having \textit{survived} for a specific follower if it has not been skipped over by that follower in irritation. To capture the intrinsic tendency of a follower $j$ to be irritated, we define a parameter $\delta_j \in [0,1]$. $\delta_j \in \{0,1\}$ correspond to always and never skipping clusters, respectively. We define a \textit{cluster survival function} $C(x;\delta_j)$ that quantifies the probability of a cluster with $x$ posts surviving follower $j$.

\begin{table}
\centering
\begin{tabular}{| l | l | l |}
    \hline
    \textbf{Model} & \textbf{Survival Function} & \textbf{Parameters} \\
    \hline
    Exponential   & $e^{-\lambda x}$             & $\lambda>0$\\
    Geometric     & $(1 - \lambda)^{x}$          & $0<\lambda \leq 1$\\
    Weibull       & $e^{-\lambda x^p}$           & $\lambda>0, p>0$\\
    Log-logistic  & $1 / (1 + \lambda x^p)$      & $\lambda>0, p>0$\\
    Rayleigh      & $e^{-x^2/2\lambda^2}$        & $\lambda>0$\\
    \hline
\end{tabular}
\caption{Common survival models, defined for $x \geq 0$.}
\label{table:survival-functions}
\end{table}

The choice of models for follower and cluster survival may depend on the specific social platform under study, or behavioural characteristics of the follower population. There are a number of models and survival functions frequently used in \textit{survival and reliability analysis}, some of which are mentioned in Table \ref{table:survival-functions}.

\hw{let's find another word instead of just `value'?} We now introduce the concept of \textit{attention potential}. For a single post at depth $d$ in the timeline to be viewed by a follower, the follower must (i) consume the timeline until depth $d$, and (ii) not skip the cluster containing this post. Concretely, a post is given attention if the follower survives until that post and the cluster containing that post survives for that follower. Since these events are independent, the probability of this happening is given by $F(d;\rho_j)C(x;\delta_j)$. This is the attention potential of a post at depth $d$ for follower $j$. The attention potential of a cluster is the sum of attention potentials of the posts within it. Given a cluster containing $x_{ij}$ posts, let there be a total of $z_{ij}$ posts on the timeline above the first post in this cluster. The attention potential of this cluster is given by the following function of the producer schedule:

\begin{equation}
\begin{aligned}
       f_{ij}(\mathbf{X}) = C(x_{ij};\delta_j)\sum_{k=1}^{x_{ij}}R(z_{ij} + k;\rho_j)
\end{aligned}
\label{eq:slot-value}
\end{equation}
\\[2pt]
The number of posts above the first post in the cluster is:

\begin{equation}
       z_{ij} = \sum_{m = 0}^{i - 1} x_{mj} + \sum_{n = 0}^{i} v_{nj}
\end{equation}
\\[2pt]
What remains is to derive the number of producer posts $x_{ij}$ and competitor posts $v_{ij}$ in the cluster from the producer schedule $\mathbf{X}$ and aggregate competitor schedules $\mathbf{C_j}$ respectively. The following relations hold:

\begin{align}
       x_{ij} &= x_{((\sigma_j - i)\bmod{S})} \label{eq:nomenclature-x}\\
       v_{ij} &= c_{((\sigma_j - i)\bmod{S})j} \label{eq:nomenclature-v}
\end{align}
\\[2pt]
Where $\bmod$ is the binary modulus operator that returns the remainder when its second argument is divided by the first.

The attention potential of the timeline constructed for a follower $j$ is the sum of attention potentials of all the producer's clusters, $\sum_{i = 0}^{S-1} f_{ij}(\mathbf{X})$. Each timeline may be weighted by a factor $\gamma_j$ encoding a preference for specific characteristics, such as influence or tendency to retweet. The attention potential of a schedule is the sum of attention potentials of all timelines:

\begin{equation}
  F(\mathbf{X}) = \sum_{i=0}^{S-1} \sum_{j=0}^{U-1} f_{ij}(\mathbf{X})
\end{equation}
\\[2pt]
The attention potential objective captures two forms of balance. The first is, for a single follower, the balance between annoying and attracting that follower's attention. The second balance is between followers with varied behavioural characteristics and activity times. An optimal schedule maximises attention potential by finding the perfect post timing and frequency configuration satisfying both these balances. We can succinctly write this optimisation problem as the following nonlinear integer program:

\begin{equation*}
\begin{aligned}
        \underset{\mathbf{X}\quad}{\text{maximise\quad}}& F(\mathbf{X})\\
        \text{subject to\quad}& \sum_{x_i \in \mathbf{X}} x_i \leq N, \\
                               & \mathbf{X} \subseteq \mathbb{Z}^S_+.\\
\end{aligned}
\end{equation*}
\\[2pt]
The form of the program reveals that it is an instance of the nonlinear knapsack problem \cite{bretthauer2002nonlinear}. Our specific objective function is \textit{non-separable}: it cannot be decomposed into a linear combination of functions $g_i(x_i)$, separate for each dimension. Additionally, it is nonconcave in general over the reals. This places the problem outside the realm for which efficient general algorithms have been devised.

Greedily exploring the discrete state space is a common local optimisation strategy. For constrained integer programs, greedy algorithms are typically applications of the method of \textit{marginal allocation}. When adapted to our problem, this corresponds to the following algorithm:

\begin{algorithm}[h]\small
       \DontPrintSemicolon
       \KwIn{Constraint $N$, initial solution $\mathbf{X}^0 \in \mathbb{Z}^S_+$}
       \Begin{
              $\mathbf{Y} \longleftarrow \mathbf{X}^0$\;
              $\mathrm{\Delta} \longleftarrow 0$\;
              \Repeat{$\Delta \leq 0$ or $\sum_{y \in \mathbf{Y}} y > N$}{
                     $i \longleftarrow \argmax_k~F(\mathbf{Y} + \mathbf{e}_k) - F(\mathbf{Y})$\;
                     $\mathbf{Y}_{prev} \longleftarrow \mathbf{Y}$\;
                     $\mathbf{Y} \longleftarrow \mathbf{Y} + \mathbf{e}_i$\;
                     $\mathrm{\Delta} \longleftarrow \mathbf{Y} - \mathbf{Y}_{prev}$\;
              }
              \Return{$\mathbf{Y}_{prev}$}
       }
       \caption{Marginal Allocation}
\end{algorithm}

Here, $\mathbf{e}_k \in \mathbb{Z}^S_+$ is the $k^{th}$ unit vector. In each iteration, the algorithm adds a single post to the slot which results in the maximum increase in the objective function value. It terminates when adding a post to any slot either violates the total intended posts constraint or does not improve the objective function value. Essentially, this is a hill-climbing algorithm starting from the initial solution and is guaranteed to terminate at an optimum, which may be local.


\section{Experiments}

In practice, social media marketers use a number of popular scheduling heuristics\footnote{https://archive.today/ykNb6} designed to optimise for attention. Using a real-world dataset from Twitter, we analyse the impact of these heuristics on the attention potential they obtain for a specific user in the dataset. Through our analyses, we discover an alternate scheduling strategy that obtains a comparatively greater attention potential while broadcasting fewer posts than the heuristic strategies.

We first select a producer whose scheduling options will be the focus of this section. We consider users with a large number of followers in the \textit{Twitter-Friends} dataset described earlier (\S2.3) in order to obtain a follower population exhibiting diverse behavioural characteristics. We also prefer users with a large number of tweets, which will aid in the behavioural parameter estimation described later in this section. We found a single user with both a large number of followers and tweets, who we adopt as the producer.

We extract the 2-hop neighbourhood in the social graph starting from the producer and obtain a subgraph with 450 followers and 813 unique competitors (other users followed by the followers). We collect all the tweets in the dataset by these users, and set $S = 24$ to construct hourly schedules.

\begin{figure}
\centering
    \subfloat[]{
        \includegraphics[width=0.23\textwidth]{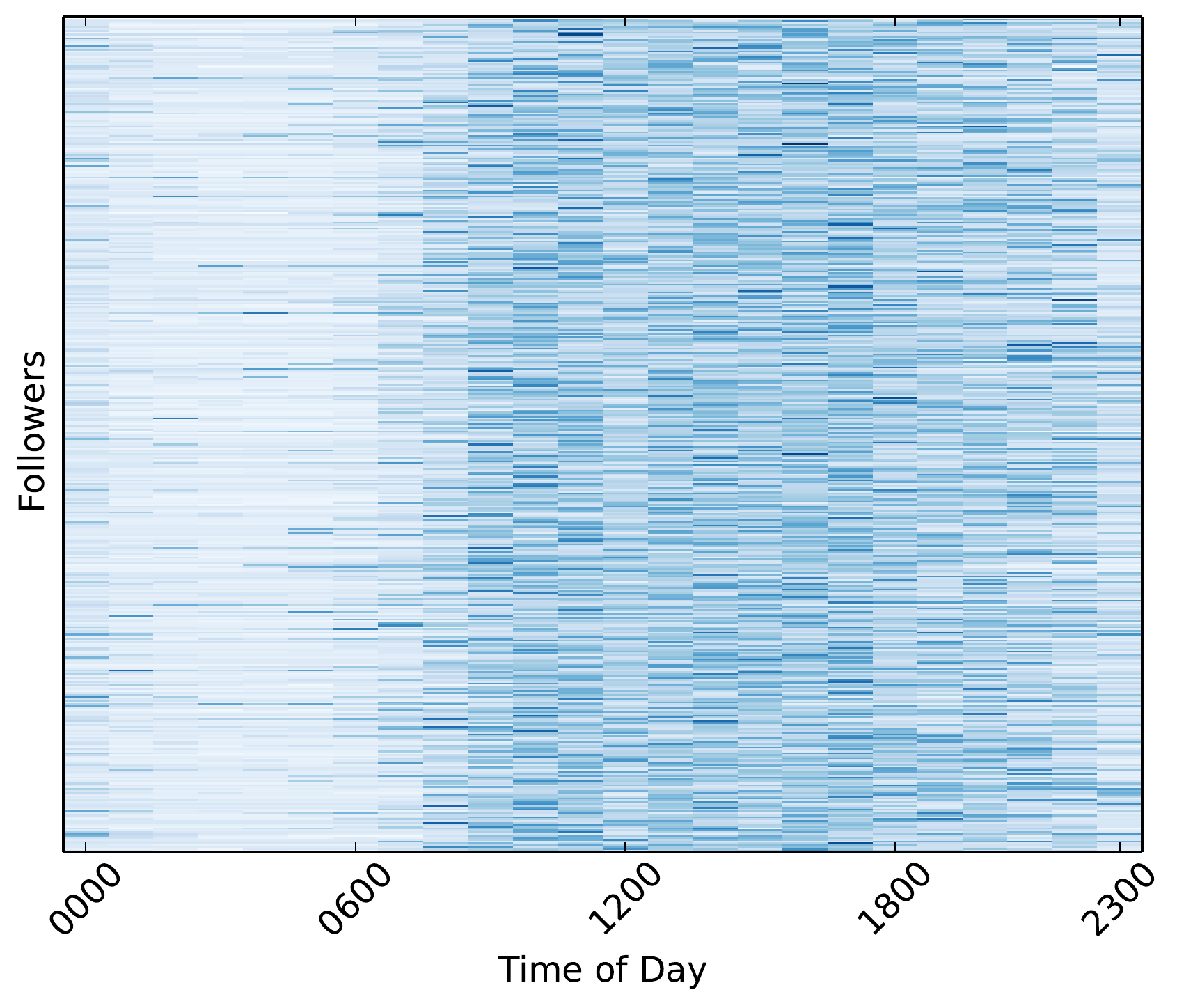}
    }
    \subfloat[]{
        \includegraphics[width=0.23\textwidth]{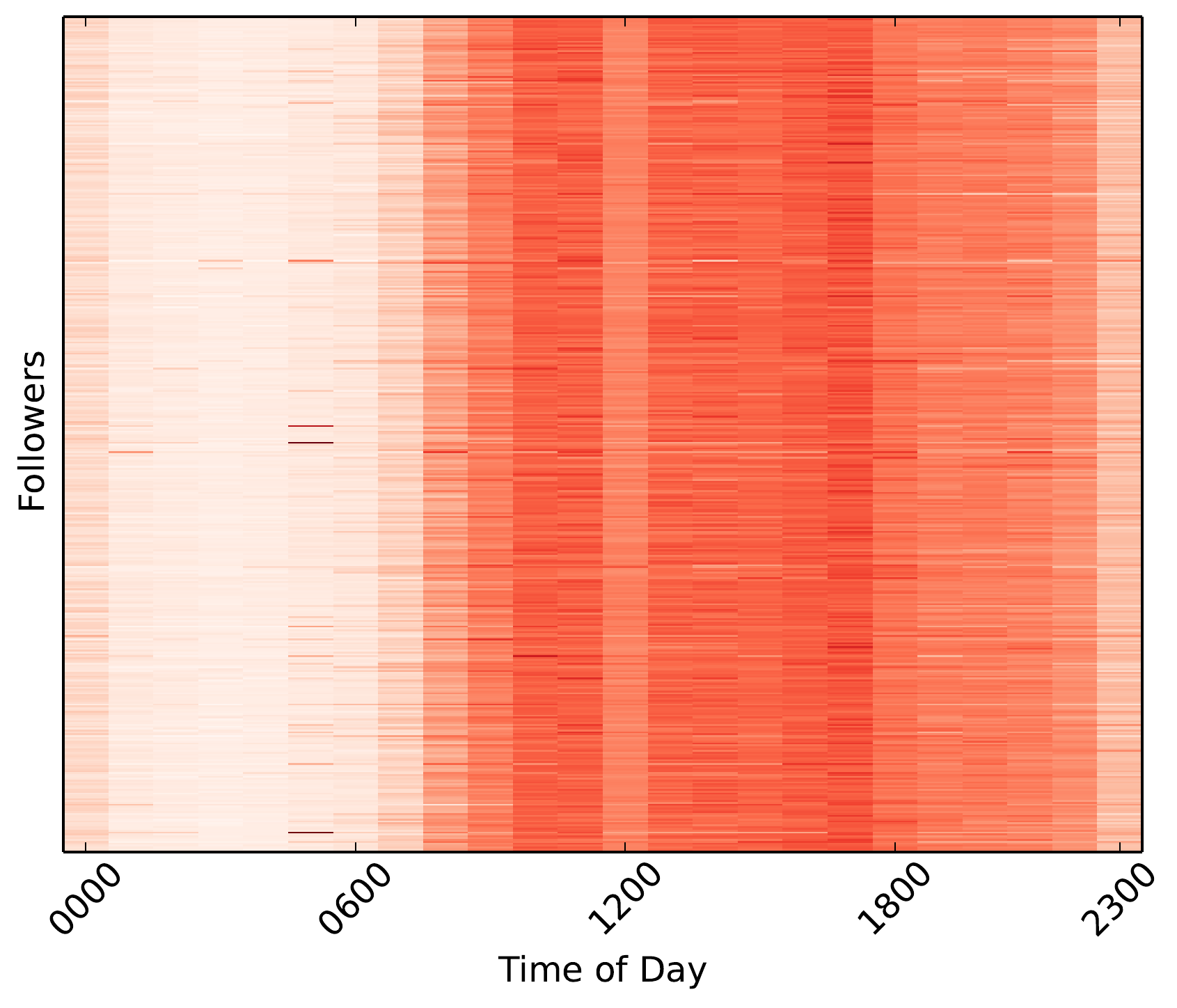}
    }
    \caption{Hourly Twitter activity of the (a) followers and (b) competitors of the producer. Times are in Pacific Standard Time (PST).}
    \label{fig:follower-competitor}
\end{figure}

To understand the collective behaviour of this user population, we observe the daily tweeting activity of the followers (Figure \ref{fig:follower-competitor}(a)) and competitors (Figure \ref{fig:follower-competitor}(b)) as a 2D histogram with hourly bins, mean-centered along each row. In both histograms, we see the most Twitter activity between 7AM and 5PM, a span of reduced activity between 6PM and 10PM and close to no activity between 11PM and 6AM. Interestingly, there is a dip in activity at 12PM, around lunchtime, which is distinctly visible in the competitor activity histogram. In fact, this has been recommended as one of the two best times to tweet\footnote{https://archive.today/rsGXm}, attributed to people using Twitter after lunch and while commuting. Our histograms suggest that reduced competition may also be a factor contributing to the increased attention obtained during these time slots.

We now estimate the behavioural parameters for each follower. We can directly estimate the login time slot $\sigma$ of a follower as the median \textit{start time slot} across all days, where a start time slot is one in which a tweet has been posted after an inactive period of over 8 hours. The 8-hour assumption is supported by the local maximum observed in the intertweet time distribution on Weibo (\S2.2).

We adopt geometric cluster and user survival functions and need to estimate their parameters from the data available. Since each user's intrinsic degrees of monotony aversion and information overload are not directly available, we use the following surrogates. We estimate the degree of monotony aversion $\delta$ of a follower towards the producer as the weakness of the follower's tie with the producer. This is supported by the finding that weaker ties induce more irritation \cite{koroleva2011cognition}. We measure tie strength as the empirical probability of retweeting or replying to the producer.

Since our user survival function arises from a geometric distribution, we can consider its parameter $\rho$ as the probability that the user stops consuming the timeline after reading a post. If we empirically calculate the average number of posts consumed by the user on each login as $\mu$, the estimate for $\rho$ is given by $1 / (1 + \mu)$.

Having estimated the followers' behavioural parameters and competitors' tweeting activity, we are now in a position to calculate the attention potential for a schedule. We evaluate the attention potentials of three scheduling heuristics along with a strategy that was surfaced from multiple runs of marginal allocation with various arbitrary initial schedules.

The \textsc{uniform} strategy advocates a fixed number of posts in each time slot. This is often recommended in rudimentary infographics discussing the optimal post frequency for social media marketers. In the \textsc{peak} strategy, posts are scheduled during the times of peak follower activity. The \textsc{graveyard} strategy is the complement of \textsc{peak}; posts are scheduled in the late night and early morning hours; often termed the ``graveyard shift''. This is a strategy recommended to take advantage of the \textit{late-night infomercial effect}\footnotemark[5]. The \textsc{smart} strategy schedules posts during lunchtime and late night hours. Both the \textsc{peak} and \textsc{smart} strategies prefer times of reduced competition.

\begin{figure}
    \includegraphics[width=0.45\textwidth]{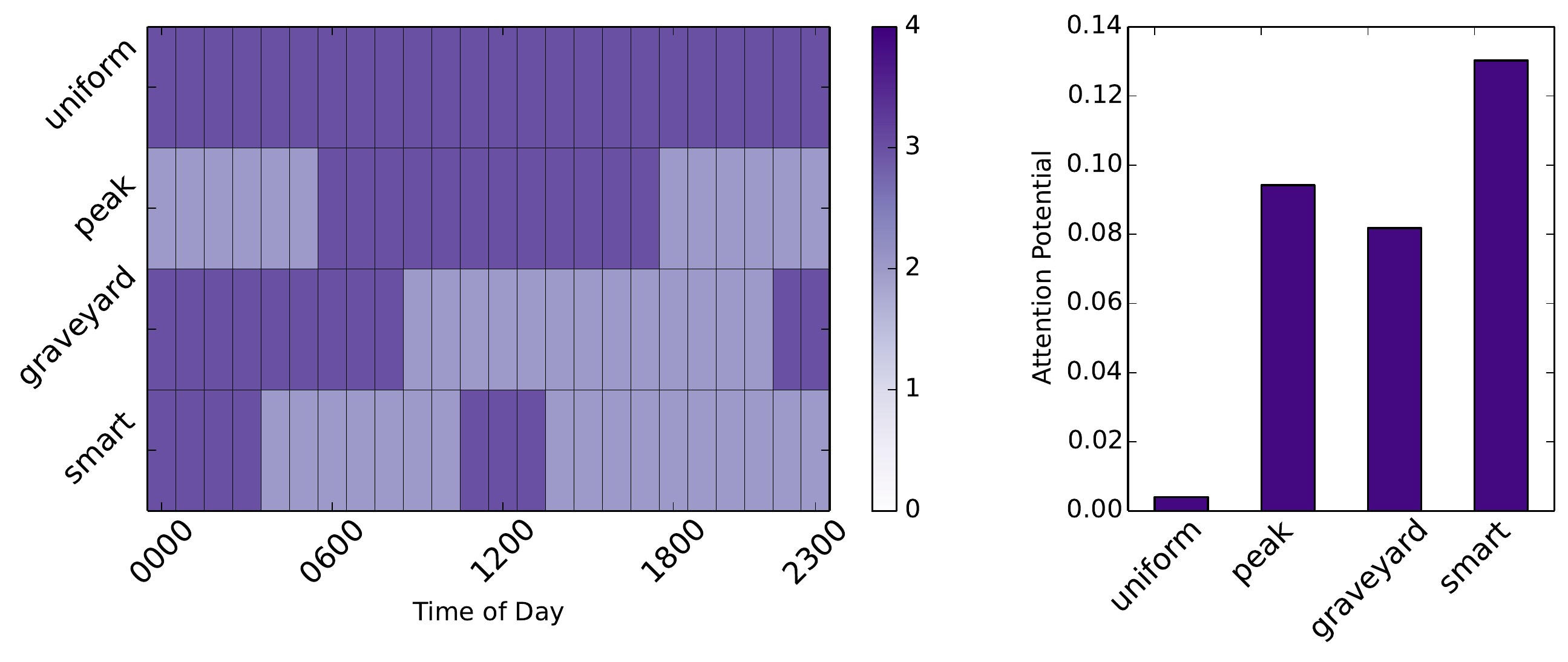}
    \caption{Post frequencies of 4 scheduling strategies and their total attention potential.}
    \label{fig:performance}
\end{figure}

The post frequencies in each time slot and total attention potential of each schedule are depicted in Figure \ref{fig:performance}. Rather counter-intuitively, the \textsc{smart} strategy had largest attention potential with the \textit{least number of posts}. The \textsc{peak} and \textsc{graveyard} strategies had similar attention potentials, indicating a close tie between optimising for follower activity and reduced competition. The \textsc{uniform} strategy had the least attention potential, possibly due to users being annoyed by the constant stream of posts.

\begin{figure*}
\centering
    \subfloat[]{
        \includegraphics[width=0.23\textwidth]{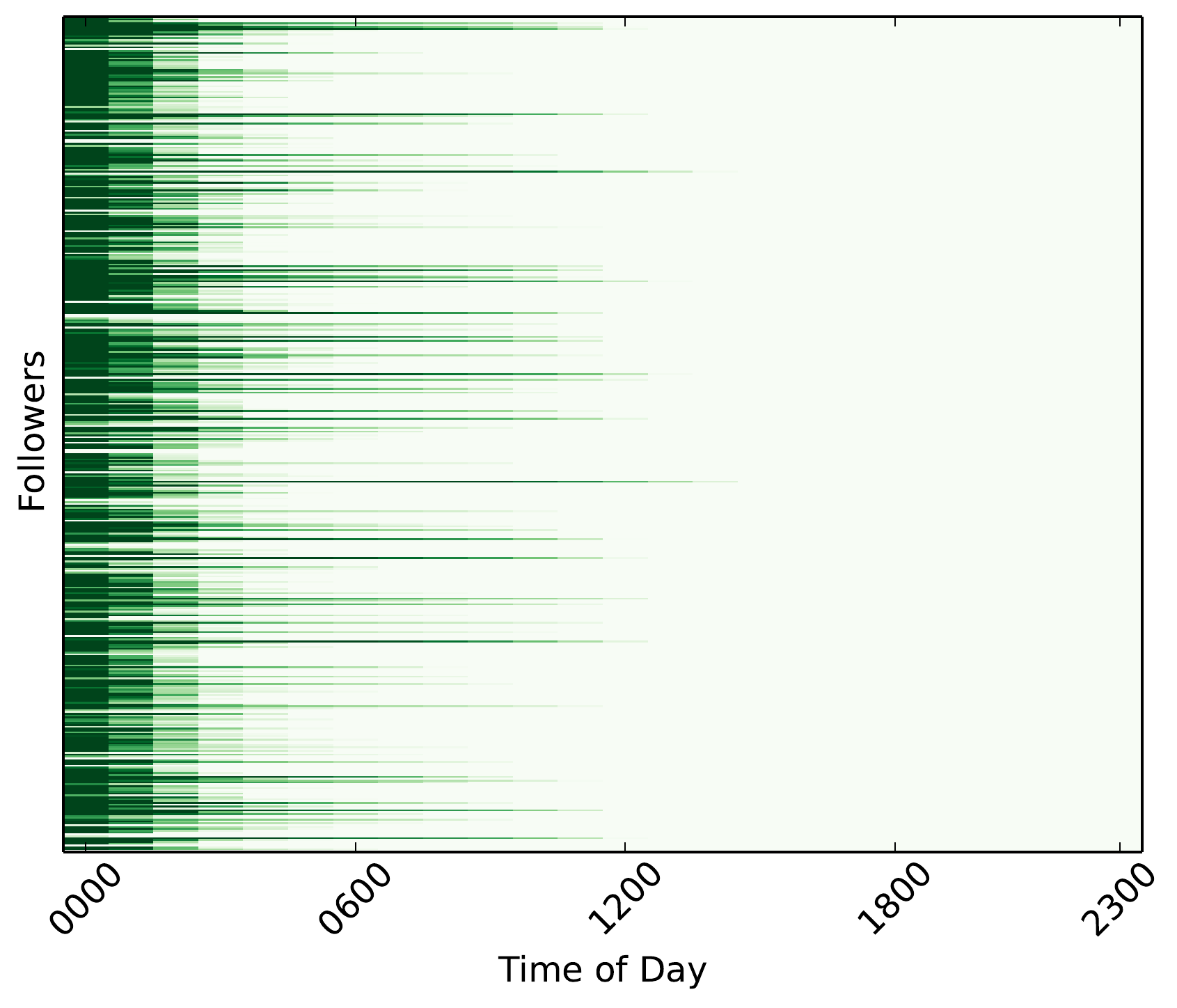}
        \label{fig:experiment-uniform}
    }
    \subfloat[]{
        \includegraphics[width=0.23\textwidth]{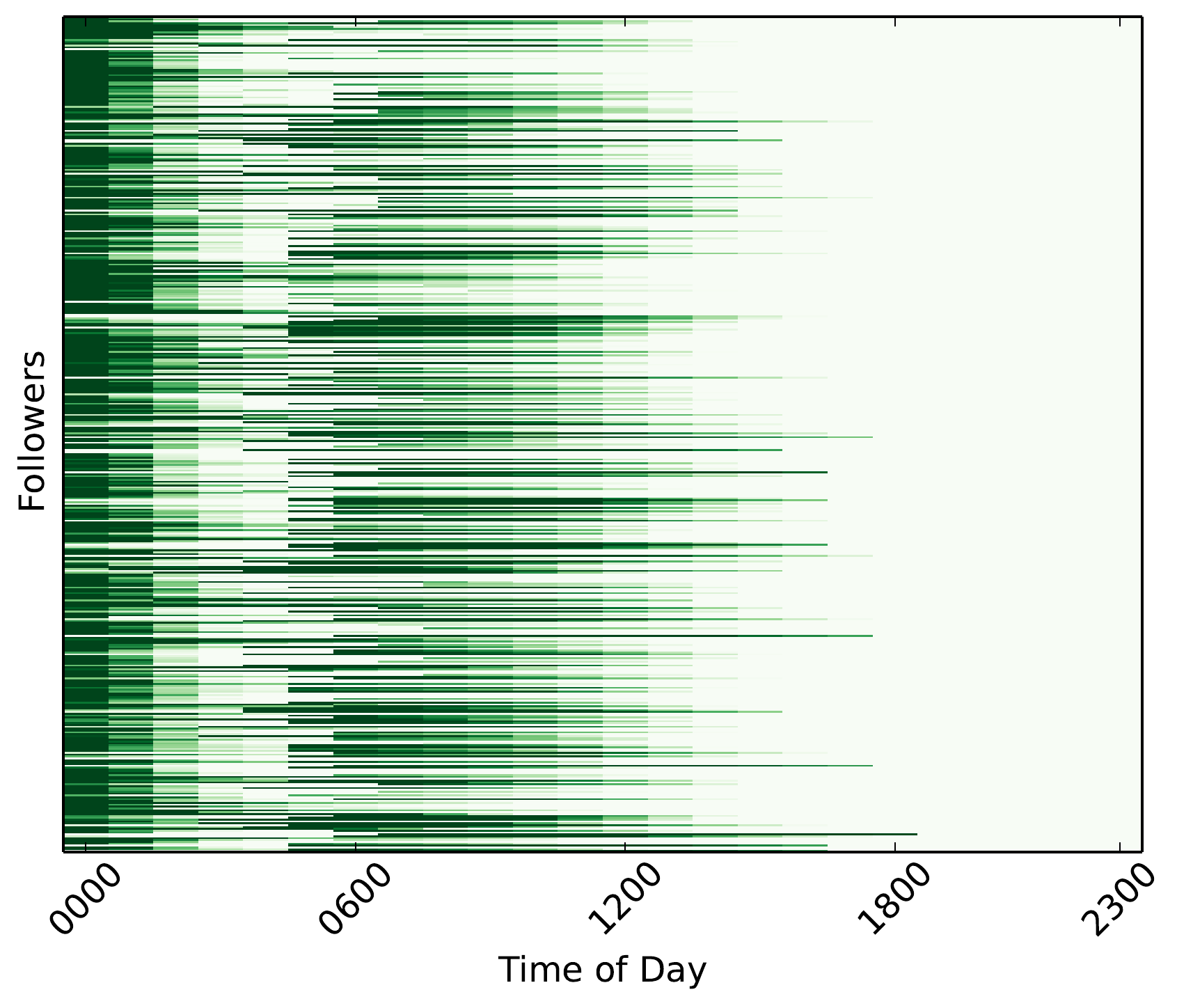}
        \label{fig:experiment-peak}
    }
    \subfloat[]{
        \includegraphics[width=0.23\textwidth]{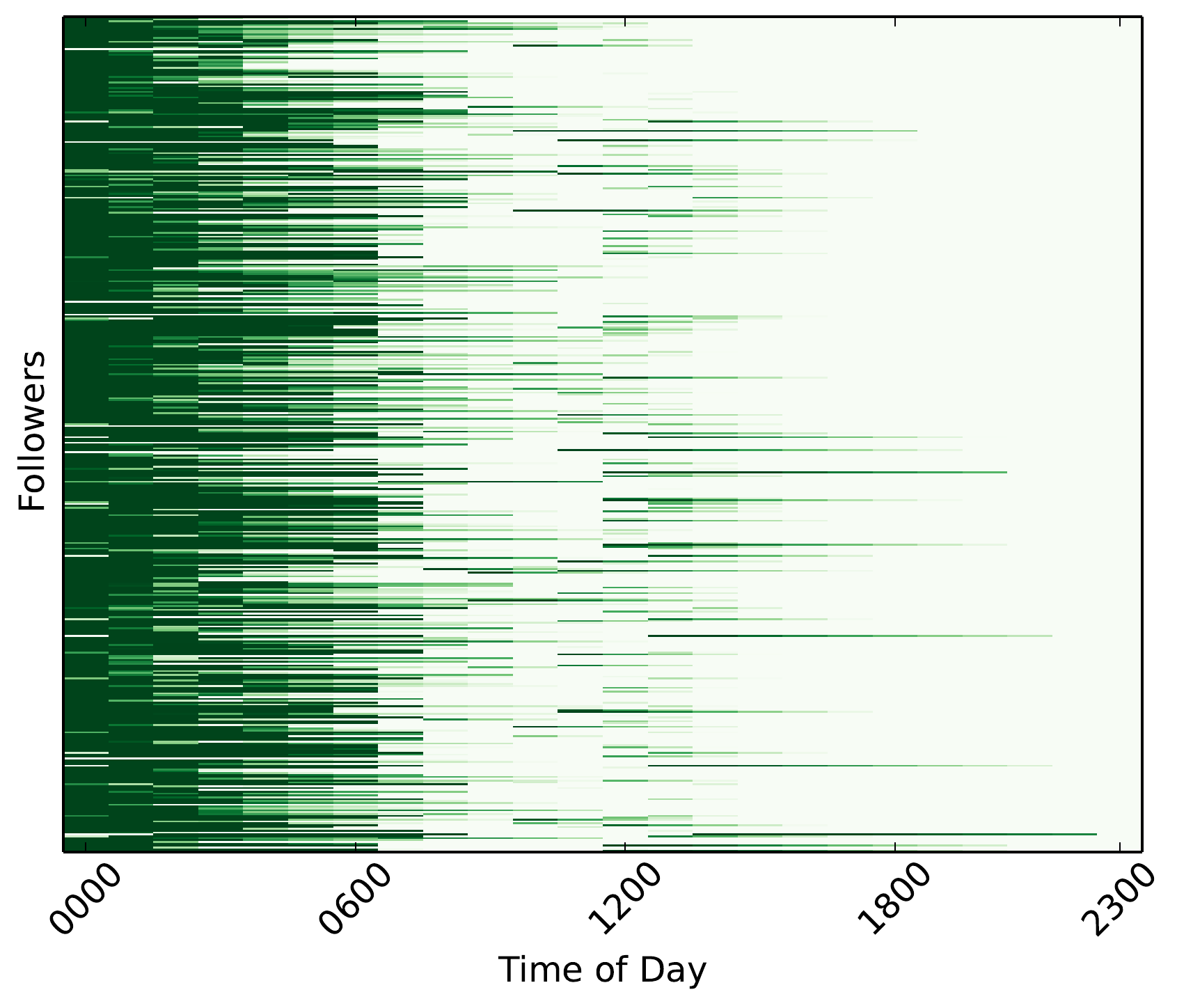}
        \label{fig:experiment-graveyard}
    }
    \subfloat[]{
        \includegraphics[width=0.23\textwidth]{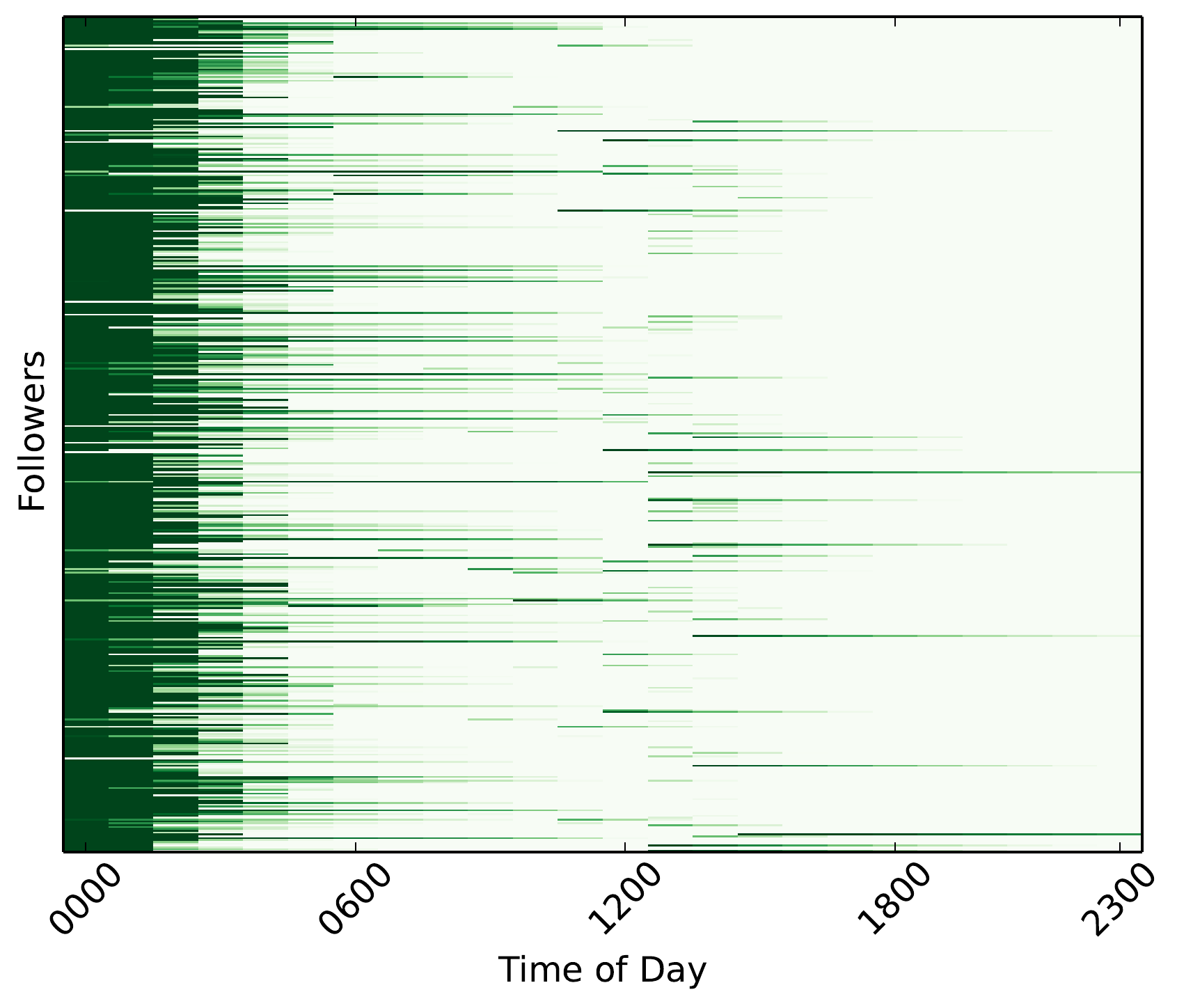}
        \label{fig:experiment-intelligent}
    }
    \label{fig:attention-potential-comparision}
    \caption{Heatmaps of the attention potential obtained when posts are scheduled (a) uniformly across all times, (b) weighted towards times of peak activity, (c) weighted towards late night and early morning hours, and (d) weighted towards lunchtime, late night and early morning hours.}
\end{figure*}

To gain further insight, we observe the attention potential histograms. mean-centered along rows, for each of these strategies in Figure 9. Some interesting patterns emerge. Though tweets were broadcast throughout the day in the \textsc{uniform} strategy, they had the most attention potential only during the late hours of the night. Similarly, the \textsc{peak} strategy had the most attention potential in the late night and early morning hours, though most posts were published during the hours of peak follower activity. Competition appears to be a dominating influence on the attention attainable by this producer.

The value of broadcast scheduling is evident when comparing the attention potential histograms for the \textsc{graveyard} and \textsc{smart} strategies. Both strategies avoid competition. However, \textsc{graveyard} publishes relatively more posts, all of which are localised around the late night and early morning hours. \textsc{smart} distributes posts between the late night and lunchtime hours, with fewer posts overall than \textsc{graveyard}. Even with fewer posts, \textsc{smart} has a greater attention potential. This may be attributed to the careful selection of time slots to maximise attention, while keeping post frequency low to avoid irritation.

In summary, the experiments in this section demonstrate the potential impact of scheduling in an attention marketplace. They also demonstrate the process of applying our scheduling framework, including the estimation of behavioural parameters from observable surrogates, to gain insights into the suitability of different scheduling strategies for the target follower population.

\pagebreak
\section{Related Work}

Broadcast scheduling techniques have been studied in the traditional attention marketplaces of television and radio programming \cite{eastman2012media,vane1994programming}. A rich assortment of strategies has evolved to deal with the problem of retaining attention and maximising advertising revenue in the face of competition. The schedules derived from these strategies often stretch over months or years, but have limited feedback on the actual attention they capture. Only recently has viewing data from set-top boxes been used to better model users and compute metrics such as audience externalities \cite{wilbur2013audience}.

The advent of the web brought us fine-grained traces of user attention and satisfaction in the form of clicks. This has enabled much work on the problem of maintaining the \textit{diversity} and \textit{novelty} of results returned by recommender systems \cite{Ziegler:2005:IRL:1060745.1060754} and search engines \cite{Agrawal:2009:DSR:1498759.1498766}. The desire for novelty has been studied extensively in the consumer research domain \cite{Kahn1995139} and recently been mined and quantified in a number of heterogeneous domains online \cite{Zhang:2014:MNT:2566486.2567976}. Maintaining novelty may conflict with the goal of maximising result accuracy, but surprisingly improves user satisfaction \cite{McNee:2006:AEA:1125451.1125659}. This aligns with our empirical findings on monotony aversion in timelines, suggesting similar roots in consumer psychology.

A related subject of recent study is the \textit{reconsumption} of items after a gap of \textit{satiation} or \textit{boredom} \cite{Kapoor:2015:JTR:2684822.2685306,Anderson:2014:DRC:2566486.2568018}, which explores the temporal cycle of the perceived novelty of an item. The focus of the task is on predicting whether and after what time gap an item is reconsumed. Our scheduling problem can be considered a variant of this task that focuses on control, where the time gap between posts is adjusted to minimise boredom and maximise reconsumption.

The problem of \textsc{Stream-Advertising} \cite{Ieong:2014:AS:2566486.2568030} is closely related. It considers a geometric decay in the reward of an advertisement with depth in the timeline, and formulates the problem as that of matching ads to slots on the timeline to maximise the total reward. The formulation explicitly prohibits consecutive ads on the timeline and ignores user irritation. Introducing our quantification of monotony into the \textsc{Stream-Advertising} problem and validating the model’s performance on real-world data could be an interesting line of future work.

Directly related to us is the problem of timing tweets to increase information campaign effectiveness \cite{dabeer2011timing}. This work is the first to formalise the tweet timing problem, frame it as a Markov decision process and present a few offline algorithms to solve it. It also discusses user irritation. However, it restricts focus to the binary decision of whether to tweet in a given time slot. Hence, it fails to address the post frequency problem. Irritation is also a considered a discrete state of the user, which increments in unit steps for each tweet by the producer that is not responded to. This model of irritation is unintuitive and remains unvalidated.

Tangentially relevant are a series of works that consider competition over timelines through the lens of game theory \cite{altman2014strategic,reiffers2014game}. These works decompose user behaviour into random processes, formulate timeline information exchange as a game and study its equilibria. Numerical results are presented with some real data being used for parameter estimation. The focus is heavily on analysis and not on control, and hence remains disconnected from the scheduling problem.

\section{Conclusions and Future Work}

We have formulated the broadcast scheduling problem that specifies when and how much to broadcast in a social network with timelines, given a set of followers and competitors with varying behavioural characteristics. The process of timeline information exchange formulated here combines user behavioural phenomena and quantifies their connection with a post's position on the timeline and the attention it is expected to receive. We also introduce and quantify monotony aversion for the first time, noting its influence in a number of diverse domains. We present experiments on data from Twitter analysing the attention potential of a number of heuristic schedules, and discover an alternate schedule via marginal allocation. This schedule outperforms the heuristics with fewer posts and has a sound intuitive basis, which is to prefer reduced competition while also limiting the amount of irritation induced.

Being the first work on this problem, there are a number of lines directly related to the problem along which progress could be fruitful. One of these is to extend the timeline information exchange process to accommodate multiple login times; this will effectively introduce more than one timeline for each user, corresponding to each login time slot. A continuous formulation in the form of random processes to replace discrete time slots is another direction worth pursuing. Extending the objective to be the ``organic reach'' by considering the spread of posts through the network may also be of interest.

While indirectly related, further study on monotony aversion is an exciting direction; there is room for a well-founded connection to be made between offline consumer behaviour, traditionally observed in physical markets, and online behaviour during the consumption of information. The differences are as important as the similarities: information arrives at a much faster pace than product purchase decisions, which may lead to significant behavioural differences. Additionally, there is no work quantifying the relation between monotony aversion and breaking ties, though a connection has been conjectured \cite{kwak2011fragile}.


%
\bibliographystyle{abbrv}

\bibliography{sigproc}  

\begin{thebibliography}{10}

\bibitem{Agrawal:2009:DSR:1498759.1498766}
R.~Agrawal, S.~Gollapudi, A.~Halverson, and S.~Ieong.
\newblock Diversifying search results.
\newblock WSDM '09.

\bibitem{alstott2014powerlaw}
J.~Alstott, E.~Bullmore, and D.~Plenz.
\newblock powerlaw: A python package for analysis of heavy-tailed
  distributions.
\newblock {\em PLoS ONE}, 9(1), 01.

\bibitem{altman2014strategic}
E.~Altman and N.~Shimkin.
\newblock {Strategic posting times over a shared publication medium}.
\newblock NetGCOOP ’14.

\bibitem{Anderson:2014:DRC:2566486.2568018}
A.~Anderson, R.~Kumar, A.~Tomkins, and S.~Vassilvitskii.
\newblock The dynamics of repeat consumption.
\newblock WWW '14.

\bibitem{backstrom2011center}
L.~Backstrom, E.~Bakshy, J.~M. Kleinberg, T.~M. Lento, and I.~Rosenn.
\newblock Center of attention: How facebook users allocate attention across
  friends.
\newblock ICWSM '11.

\bibitem{Bernstein:2013:QIA:2470654.2470658}
M.~S. Bernstein, E.~Bakshy, M.~Burke, and B.~Karrer.
\newblock Quantifying the invisible audience in social networks.
\newblock CHI '13.

\bibitem{Bollen:2010:UCO:1864708.1864724}
D.~Bollen, B.~P. Knijnenburg, M.~C. Willemsen, and M.~Graus.
\newblock Understanding choice overload in recommender systems.
\newblock RecSys '10.

\bibitem{bretthauer2002nonlinear}
K.~M. Bretthauer and B.~Shetty.
\newblock The nonlinear knapsack problem--algorithms and applications.
\newblock {\em European Journal of Operational Research}, 138(3):459--472,
  2002.

\bibitem{chun2008comparison}
H.~Chun, H.~Kwak, Y.-H. Eom, Y.-Y. Ahn, S.~Moon, and H.~Jeong.
\newblock Comparison of online social relations in volume vs interaction: a
  case study of cyworld.
\newblock SIGCOMM '08.

\bibitem{comarela2012understanding}
G.~Comarela, M.~Crovella, V.~Almeida, and F.~Benevenuto.
\newblock Understanding factors that affect response rates in twitter.
\newblock HT '12.

\bibitem{counts2011taking}
S.~Counts and K.~Fisher.
\newblock Taking it all in? visual attention in microblog consumption.
\newblock ICWSM '11.

\bibitem{dabeer2011timing}
O.~Dabeer, P.~Mehendale, A.~Karnik, and A.~Saroop.
\newblock Timing tweets to increase effectiveness of information campaigns.
\newblock ICWSM '11.

\bibitem{eastman2012media}
S.~Eastman and D.~Ferguson.
\newblock {\em Media programming: Strategies and practices}.
\newblock Cengage Learning, 2012.

\bibitem{eppler2004concept}
M.~J. Eppler and J.~Mengis.
\newblock The concept of information overload: A review of literature from
  organization science, accounting, marketing, mis, and related disciplines.
\newblock {\em The information society}, 20(5):325--344, 2004.

\bibitem{fu2013assessing}
K.-w. Fu, C.-h. Chan, and M.~Chau.
\newblock Assessing censorship on microblogs in china: discriminatory keyword
  analysis and the real-name registration policy.
\newblock Internet Computing '13.

\bibitem{golder2011diurnal}
S.~A. Golder and M.~W. Macy.
\newblock Diurnal and seasonal mood vary with work, sleep, and daylength across
  diverse cultures.
\newblock {\em Science}, 333(6051), 2011.

\bibitem{gomez2014quantifying}
M.~Gomez-Rodriguez, K.~P. Gummadi, and B.~Sch{\"o}lkopf.
\newblock Quantifying information overload in social media and its impact on
  social contagions.
\newblock ICWSM '14.

\bibitem{gonccalves2011modeling}
B.~Gon{\c{c}}alves, N.~Perra, and A.~Vespignani.
\newblock Modeling users' activity on twitter networks: Validation of dunbar's
  number.
\newblock {\em PloS One}, 2011.

\bibitem{Ieong:2014:AS:2566486.2568030}
S.~Ieong, M.~Mahdian, and S.~Vassilvitskii.
\newblock Advertising in a stream.
\newblock WWW '14.

\bibitem{jo2012circadian}
H.-H. Jo, M.~Karsai, J.~Kert{\'e}sz, and K.~Kaski.
\newblock Circadian pattern and burstiness in mobile phone communication.
\newblock {\em New Journal of Physics}, 2012.

\bibitem{Kahn1995139}
B.~E. Kahn.
\newblock Consumer variety-seeking among goods and services: An integrative
  review.
\newblock {\em Journal of Retailing and Consumer Services}, 2(3), 1995.

\bibitem{Kapoor:2015:JTR:2684822.2685306}
K.~Kapoor, K.~Subbian, J.~Srivastava, and P.~Schrater.
\newblock Just in time recommendations: Modeling the dynamics of boredom in
  activity streams.
\newblock WSDM '15.

\bibitem{kim2013microscopic}
J.~Kim, D.~Lee, and B.~Kahng.
\newblock Microscopic modelling circadian and bursty pattern of human
  activities.
\newblock {\em PloS one}, 2013.

\bibitem{koroleva2011cognition}
K.~Koroleva, H.~Krasnova, and O.~G{\"u}nther.
\newblock {\em Cognition or affect? Exploring information processing on
  Facebook}.
\newblock Springer, 2011.

\bibitem{kwak2011fragile}
H.~Kwak, H.~Chun, and S.~Moon.
\newblock Fragile online relationship: a first look at unfollow dynamics in
  twitter.
\newblock CHI '11.

\bibitem{lin2014steering}
S.~Lin, Q.~Hu, F.~Wang, and P.~S. Yu.
\newblock Steering information diffusion dynamically against user attention
  limitation.
\newblock ICDM '14.

\bibitem{McNee:2006:AEA:1125451.1125659}
S.~M. McNee, J.~Riedl, and J.~A. Konstan.
\newblock Being accurate is not enough: How accuracy metrics have hurt
  recommender systems.
\newblock CHI EA '06.

\bibitem{mryglod2015interevent}
O.~Mryglod, B.~Fuchs, M.~Szell, Y.~Holovatch, and S.~Thurner.
\newblock Interevent time distributions of human multi-level activity in a
  virtual world.
\newblock {\em Physica A: Statistical Mechanics and its Applications}, 2015.

\bibitem{reiffers2014game}
A.~Reiffers-Masson, E.~Altman, and Y.~Hayel.
\newblock A time and space routing game model applied to visibility competition
  on online social networks.
\newblock NETGCOOP '14.

\bibitem{Shahaf:2010:CDN:1835804.1835884}
D.~Shahaf and C.~Guestrin.
\newblock Connecting the dots between news articles.
\newblock KDD '10.

\bibitem{vane1994programming}
E.~T. Vane.
\newblock {\em Programming for TV, radio, and cable}.
\newblock Taylor \& Francis US, 1994.

\bibitem{wilbur2013audience}
K.~C. Wilbur, L.~Xu, and D.~Kempe.
\newblock Correcting audience externalities in television advertising.
\newblock {\em Marketing Science}, 32(6), Nov 2013.

\bibitem{yasseri2012circadian}
T.~Yasseri, R.~Sumi, and J.~Kert{\'e}sz.
\newblock Circadian patterns of wikipedia editorial activity: A demographic
  analysis.
\newblock {\em PloS One}, 2012.

\bibitem{Zhang:2014:MNT:2566486.2567976}
F.~Zhang, N.~J. Yuan, D.~Lian, and X.~Xie.
\newblock Mining novelty-seeking trait across heterogeneous domains.
\newblock WWW '14.

\bibitem{Ziegler:2005:IRL:1060745.1060754}
C.-N. Ziegler, S.~M. McNee, J.~A. Konstan, and G.~Lausen.
\newblock Improving recommendation lists through topic diversification.
\newblock WWW '05.

\end{thebibliography}
\end{document}